\documentclass[showpacs,amsmath,amssymb,aps,prb,twocolumn]{revtex4}

\usepackage{bm}
\usepackage{color}

\usepackage{graphicx}

\usepackage{dcolumn}

\usepackage{natbib} 

\usepackage{hyperref}
\usepackage{pifont}

\newcommand{\et}{$et\:al.\:$}
\newcommand{\SiN}{Si$_{3}$N$_{4}\,$}

\newcommand{\ef}{$E_\mathrm{F}$}

\begin{document}


\title{First-principles study of high conductance DNA sequencing with carbon nanotube electrodes}

\author{X. Chen$^\mathrm{1}$, I.~Rungger$^\mathrm{1}$, C.D.~Pemmaraju$^\mathrm{1}$, U. Schwingenschl{\"o}gl$^\mathrm{2}$ and S. Sanvito$^\mathrm{1}$}
\affiliation{1) School of Physics and CRANN,Trinity College, Dublin 2, Ireland \\
2) PSE Division, KAUST, Thuwal 23955-6900, Kingdom of Saudi Arabia}

\date{\today}


\begin{abstract}

Rapid and cost-effective DNA sequencing at the single nucleotide level might be achieved by measuring a transverse electronic
current as single-stranded DNA is pulled through a nano-sized pore. In order to enhance the electronic coupling between the
nucleotides and the electrodes and hence the current signals, we employ a pair of single-walled close-ended (6,6) carbon nanotubes (CNTs) 
as electrodes. We then investigate the electron transport properties of nucleotides sandwiched between such electrodes by using 
first-principles quantum transport theory. In particular we consider the extreme case where the separation between the electrodes
is the smallest possible that still allows the DNA translocation. The benzene-like ring at the end cap of the CNT can strongly couple 
with the nucleobases and therefore both reduce conformational fluctuations and significantly improve the conductance. As such, when 
the electrodes are closely spaced, the nucleobases can pass through only with their base plane parallel to the plane of CNT 
end caps. The optimal molecular configurations, at which the nucleotides strongly couple to the CNTs, and which yield the largest 
transmission, are first identified. These correspond approximately to the lowest energy configurations. Then the electronic 
structures and the electron transport  of these optimal configurations are analyzed. 
The typical tunneling currents are of the order of 50~nA for voltages up to 1~V. At higher bias, where resonant transport through the 
molecular states is possible, the current is of the order of several $\mu$A. Below 1~V the currents associated to the different nucleotides 
are consistently distinguishable, with adenine having the largest current, guanine the second-largest, cytosine the third and finally 
thymine the smallest. We further calculate the transmission coefficient profiles as the nucleotides are dragged along the DNA translocation path and 
investigate the effects of configurational variations. Based on these results we propose a DNA sequencing protocol combining three 
possible data analysis strategies.

\end{abstract}


\maketitle

\clearpage
\section{\label{sec:Introduction}Introduction}

In 1996 Kasianowicz \et  demonstrated the possibility of translocating by electrophoresis single-stranded DNA (ssDNA) and RNA 
molecules through nano-sized biological channels~\citep{Kasianowicz1996}. This discovery initiated a surge of proposals for novel DNA 
sequencing protocols~\citep{Branton2008, Zwolak2008}. Among the many Zwolak \et proposed to sort the nucleobases from each other in a sequence 
by measuring the transverse tunneling current across a ssDNA as this is driven through a nanopore~\citep{Zwolak2005}. It was suggested 
that single base resolution is in principle achievable, as the differences in the electronic structure among the nucleobases may translate 
into distinguishable tunneling current signals. However, as the tunneling process is extremely sensitive to the distance between the 
electrodes and the nucleotides, one is left wondering whether the fluctuations in the transverse tunnel current, due to the fluctuations
in the local geometry, may be too large to allow the identification of the various nucleotides. Furthermore additional fluctuations may 
arise from the electrostatic gate action introduced by the solvent molecules~\cite{Ivan1,Das2010}. In order to overcome these intrinsic 
difficulties it was further suggested to trace the current fluctuations as the ssDNA is translocated across the pore, and to use their
statistical distribution in order to unambiguously recognize the electronic signatures of the various nucleobases~\citep{Lagerqvist2006, Lagerqvist2007,Krems2009}.

Experimentally, Tsutsui \et showed that it is possible to identify single nucleotides in solution by two-probe tunneling current measurements
and a thorough statistical analysis of the time-resolved current \citep{Tsutsui2010}. In this case the electrodes could be mechanically moved
to optimize the current through a given nucleotide. As such, although the experiment proves the concept of using electrical currents for
the sequencing, it does not demonstrate a working device. 

Prototype sequencing devices with electrodes integrated 
in solid-state nanopores and nanofluidic channels have been also proposed\citep{Gierhart2008, Liang2008, Jiang2010, 
Spinney2009, Ivanov2010}. These have been realized experimentally by high resolution milling-based methods for a number
of metal electrodes, but it has been speculated that similar techniques could be used for CNTs with possibly higher resolution
\citep{Jiang2010}. On the theoretical front, the transverse tunneling conductance across 
nucleobases placed between two gold electrodes has been actively investigated and debated \citep{Zwolak2005, Lagerqvist2006,Zikic2006, 
Lagerqvist2007comment, Zikic2007reply, Zhang2006, He2008, He2010}. Interestingly recently some special attention has been dedicated 
to exploring graphene nanopore, graphene nanoribbon and carbon nanotubes (CNT) as potential electrodes materials
\citep{Nelson2010,Prasongkit2010,Meunier2008,Nikolic2011}.

Despite these many works a key question still remains largely unanswered, namely how can one enhance the nucleotide-electrode 
interaction to a point where the transmigration is still possible, but the geometrical fluctuations are sufficiently suppressed to allow 
unambiguous single nucleotide recognition. Possible strategies for achieving this goal are based on functionalizing the electrodes 
with various chemical agents (including nucleobases themselves for example), which interact differently with the different nucelobases 
\citep{Ohshiro2006, Meunier2008, He2007, He2008}. Yet, the functionalization may be extremely challenging in particular if this
needs to be selective at the length scale of the typical nanopore (a few nm). Here we use a different strategy that does not involve
any chemical functionalization and considers single-walled close-ended (6,6) CNT electrodes placed at an ultra-short distance 
[see Fig. \ref{Fig:CNT-Device}~(a)]. An important feature of these electrodes is that the benzene-like six-membered ring at the 
closed end cap provides the possibility of $\pi$-$\pi$ coupling with the nucleobases, as ssDNA is transmigrated across. Importantly 
the $\pi$-$\pi$ channel forces the nucleotides to align flat with respect to the electrode cap so that conformal fluctuations are 
highly suppressed. Our work analyses in detail the electron transport in this favourable condition and propose a protocol for 
distinguishing the nucleotides. 

In short our computational strategy unfolds as follows. We first overview the general characteristics of the transmission coefficient 
as a function of energy for the four nucleotides sandwiched between two such electrodes. Then we search for the optimal molecular
configurations of the nucleotides about the electrodes. For these configurations the electronic coupling is maximized and so is the 
low bias current. The electronic structure, the zero-bias transmission coefficients, and the $I$-$V$ curves for all the nucleotides are 
analyzed next at the resulting optimal configurations. Finally we calculate the transmission coefficient profiles along the full DNA 
translocation path and propose a DNA sequencing protocol combining multiple data analysis. A method and a conclusion section 
complete this work.

\section{Method}
\label{sec:Method}

\begin{figure}
 \center
 \includegraphics*[width=3.25in,clip=true]{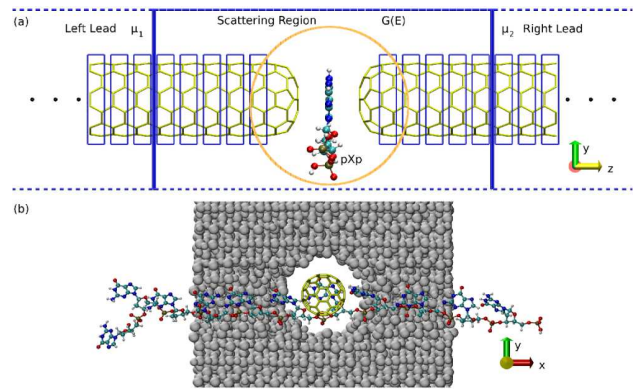}
 \caption{The device setup considered in this work. In (a) we present the CNT/pXp/CNT (X=A, C, G, and T) junction used to
 calculate the electron transport. The left- and right-hand side (6,6) CNT electrodes are semi-infinite and periodic along the transport 
 direction ($z$-axis). In the scattering region the pXp molecules are sandwiched between the CNTs. A few layers of carbon atoms 
 of both the CNTs are included in the scattering region allowing the convergence of charge density at the boundaries with the 
 semi-infinite electrodes. In (b) a typical MD snapshot of the entire device, where the CNT electrodes are embedded into a 
 \SiN-based nanopore [sketched by the circle in panel (a)], and an 11-base oligo-deoxyriboguanosine translocates through 
 the nanopore (grey, cutaway view). The single-stranded DNAs translocate through the nanopore along the $x$-axis. A guanine base 
 can be seen passing the gap between CNTs. Water and electrolyte ions are not displayed for clarity.}
 \label{Fig:CNT-Device}
\end{figure}

The device proposed here consists of a pair of semi-infinite close-ended CNT (6,6) electrodes aligned along the $z$-axis, 
and embedded inside a \SiN nanopore [see Fig.~\ref{Fig:CNT-Device}]. The separation between the CNTs' caps is initially 
only $6.6$~\AA\ (this will be then optimized as described later). Since the carbon van der Waals radius 
is about $1.7$~\AA, one may expect that the small space between the electrodes will only accommodate the planar nucleobases 
with their base plane parallel to the plane of CNT end caps, but not the non-planar sugar ring and phosphate groups. 
This is indeed confirmed by molecular dynamics (MD) simulations performed with empirical potentials for a ssDNA pulled 
through a \SiN nanopore [see the snapshot in Fig. \ref{Fig:CNT-Device}(b)], from which we find that in general the 
nucleobases pass the gap between the CNTs approximately as indicated Fig. \ref{Fig:CNT-Device}(a). A detailed description of the
MD simulations for the DNA translocation is outside the scope of the present paper and it will be published elsewhere. Here we focus 
only on the transport properties of the ideal case. We note that not only does such a setup minimize the structural fluctuations during 
translocation, but it also leads to an effective $\pi$-$\pi$ bonding and to a strong electronic coupling between the nucleotides and both 
the electrodes. We therefore expect a much larger current than that for systems where the nucleotides are oriented perpendicular to 
the electrodes, i.e. for the setup considered in several other studies\citep{He2008, Prasongkit2010, Meunier2008}.
%

The four nucleotides commonly found in DNA are represented by four neutral chemical species, namely, adenosine 3'-phosphate-5'-phosphate 
(pAp), cytidine 3'-phosphate-5'-phosphate (pCp), guanosine 3'-phosphate 5'-phosphate (pGp) and thymidine 3'-phosphate-5'-phosphate (pTp). 
The two phosphate groups at both sides of each nucleobase are included in order to approximate the nearest 
chemical environment. In the initial configurations, each of the pXp (X=A, C, G, and T) molecules is sandwiched between the two 
CNT end caps, with their nucleobase mass centers coinciding with the origin. This is defined as the middle point of the electrode 
gap and the bonds connecting the sugar ring and the nucleobases aligned along the $y$-axis. We assume that the nucleotides 
translocate along the $x$-axis. The unit cell size is set so that the distance between molecules in neighboring cells is at least 
$20$~\AA, along both the $x$ and $y$ direction (we use periodic boundary conditions in the direction orthogonal to that of the
transport, namely in the $x$-$y$ plane).

The computation of the transmission coefficients ($T$) and the current-voltage ($I$-$V$) characteristics is performed by using 
the \textit{ab initio} electronic transport code {\sc Smeagol}\cite{Rocha2005,Rocha2006,Rungger2008}. {\sc Smeagol} implements 
the non-equilibrium Green's function (NEGF) method over density functional theory (DFT), by using the pseudopotential code 
{\sc Siesta} \cite{Soler2002} as its electronic structure platform. In {\sc Smeagol} the device under investigation is partitioned into 
three regions: the left- and right-hand side semi-infinite current-voltage electrodes, and the scattering region 
[see Fig.~\ref{Fig:CNT-Device}(a)]. In our calculations the scattering region comprises 156 and 144 carbon atoms respectively
of left- and right-hand side CNT electrodes, and one of the pXp molecules. The portion of the electrodes included explicitly in 
the scattering region is large enough to ensure that the calculated charge density at the outermost layers converges to that of 
bulk the CNTs.

Both the electronic structure and the transport properties are calculated with the local density approximation (LDA) of the DFT
exchange and correlation functional. A double-$\zeta$ plus polarization basis sets is used for C($2s2p$), N($2s2p$), 
O($2s2p$), P($3s3p$) and H($1s$) throughout. All calculations are carried out with an equivalent real space mesh cutoff 
of $300$~Ry and an electronic temperature of $300$~K. In {\sc Smeagol} the charge density is evaluated by separating the 
integral over the non-equilibrium Green's function into an equilibrium part, which is performed along a contour in the complex 
energy plane, and a non-equilibrium part, which is performed along the real energy axis~\cite{Rocha2006}. For the first we use 16 
energy points on the complex semi-circle, 16 points along the line parallel to the real axis and 16 poles. The integral over real energies 
necessary at finite bias is evaluated over a mesh, whose energy spacing is not larger than 1~meV~\cite{Rocha2005,Rocha2006}.

\section{\label{sec:Results}Results and discussion}

\subsection{\label{sec:TatFermiLevel}General characteristics of the transmission at zero bias}

\begin{figure}
  \center
  \includegraphics*[width=9.0cm,clip=true]{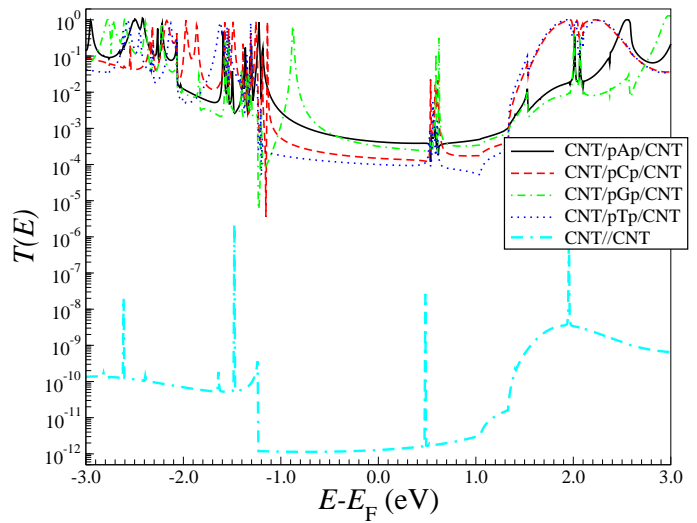}
  \caption{Zero-bias transmission coefficient as a function of energy for CNT//CNT and the four CNT/pXp/CNT (X=A, C, G, and T) 
           junctions. The curves refer to the initial configurations where the nucleobase' center of mass coincides with the middle 
           point of the electrode gap. The CNT electrodes are $6.6$~\AA\ apart.}
  \label{Fig:InitialConfig-Tat0V}
\end{figure}

We first examine the general transport properties of all the possible four CNT/pXp/CNT (X=A, C, G, and T) junctions by plotting, 
in Fig.~\ref{Fig:InitialConfig-Tat0V}, the zero-bias transmission coefficient as function of energy for the initial geometries, 
where the nucleobases' center of mass coincides with the middle point of the electrode gap. Electron tunneling between 
the CNT electrodes ($6.6$~\AA\, apart in these initial simulations) through vacuum (we denote this configuration as 
CNT//CNT) is extremely small ($T\sim1.3\times10^{-12}$) around the Fermi level, $E_\mathrm{F}$. Note that the linear
response (zero-bias) conductance, $G$, is simply $G=G_0\:T({E_\mathrm{F}})$, where $G_0=2e^2/h$ is the quantum 
conductance, $e$ is the electronic charge and $h$ the Planck constant, so that a tiny $T({E_\mathrm{F}})$ is equivalent
to a small linear response conductance. 
%

The inclusion of the pXp molecules in the device results in an increase of the transmission coefficient by over eight orders of magnitude with typical 
values of $T(E_\mathrm{F})$ of around 10$^{-4}$--10$^{-3}$. We therefore expect in general a drastic change in the measured current 
when a nucleotide passes between the CNT electrodes, regardless of the nucleotide type. As such our device is at least capable 
of distinguishing whether or not a molecule is between the two electrodes. Notably the magnitude of the transmission coefficient at 
$E_\mathrm{F}$ is found to be orders of magnitude larger than that found by using either functionalized gold 
($\sim10^{-6}$)~\cite{He2008} or graphene ($>10^{-10}$)\cite{Prasongkit2010} electrodes (in both cases
the calculations have been carried out with {\sc Smeagol} using similar computational parameters). Clearly our much larger
tunneling current originates from the short electrode-electrode separation and by the fact that the nucleobases lie flat
between the electrodes. 

\begin{figure}
  \center
  \includegraphics*[width=8.5cm,clip=true]{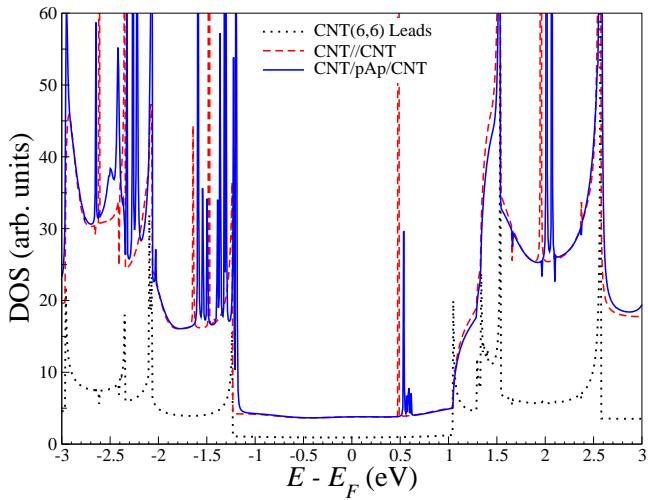}
  \caption{Density of states (DOS) of a (6,6) CNT electrode and of both the CNT//CNT and the CNT/pAp/CNT junctions
            as function of energy for the initial configurations. CNT/pAp/CNT is chosen as a representative of the CNT/pXp/CNT
            junctions.}
  \label{Fig:InitialConfig-DOS}
\end{figure}

In the various $T(E)$ profiles there are peaks, as the one clearly visible at 0.5~eV, common to all the junctions. All of these are very pronounced and 
extremely sharp and they are characteristic of the CNT//CNT device as well (see peaks at $-2.61$~eV, $-1.64$~eV, $-1.48$~eV and $1.95$~eV, with all 
the energies taken with respect to the CNT $E_\mathrm{F}$). These are all due to localized surface states 
of the close-ended CNTs and therefore are not representative of any feature of the nucleotide electronic structure. Such CNT surface 
states can be readily identified by comparing the density of states (DOS) of an infinite CNT with the one of the semi-infinite 
close-ended CNT, shown in Fig.~\ref{Fig:InitialConfig-DOS}.
%

For all the CNT/pXp/CNT junctions investigated we can identify two distinct transport regimes (Fig.~\ref{Fig:InitialConfig-Tat0V}): 
a tunneling regime for energies between $-1.0$~eV and $1.5$~eV, and a resonant transport regime through 
the nucleotides' molecular levels for energies outside this range. Around $E_\mathrm{F}$, the transmission curves for all the 
nucleotides are very smooth and vary little over energy. Importantly, the relative order of magnitude of the transmission coefficients, 
namely $T_\mathrm{pAp}$ $>$ $T_\mathrm{pGp}$ $>$ $T_\mathrm{pCp}$ $>$ $T_\mathrm{pTp}$, is preserved for a rather 
large energy range around the Fermi level. This indicates that it is possible to distinguish the nucleotides already in the 
low-bias tunneling regime, and that the recognition is relatively robust with respect to fluctuations in the position of $E_\mathrm{F}$.
Furthermore the flatness of the transmission coefficient around $E_\mathrm{F}$ suggests that one 
can infer the current for voltages up to about 1~V from the simple relation $I\approx T(E_\mathrm{F})V$. Therefore, in the next 
section we will analyze the transport properties at the Fermi energy only, and search for the molecular configurations of 
the CNT/pXp/CNT junctions that yield the largest transmission at the Fermi energy. When these configurations are explored by the ssDNA
during the translocation process they will provide the largest contribution to the current.

\subsection{\label{sec:SearchConfigs}Search for the molecular configurations yielding the maximum transmission}

For the device setup investigated here we have identified four possible degrees of freedom, which determine the relative 
position of the pXp molecules with respect to the electrodes: (1) the rotation angle of pXp around the $z$-axis [up to $60^\circ$ 
due to the (6,6) CNT symmetry], (2) a translation of the pXp molecules about the $y$- and (3) the $x$-axis and (4) a variation of the 
distance between the two CNT electrodes. Here we assume that the internal structure of the pXp molecules does not deform and 
that the nucleobases remain parallel to the CNT end cap planes during the translocation.

In section \ref{sec:ConfigutationEffect} we will discuss in details the effects of these operations on the electron transport; here we first 
identify the configurations which maximize the transmission at \ef. To this goal we start from the initial configurations of the 
CNT/pXp/CNT junctions, defined in Sec.~\ref{sec:Method}, and perform all the possible independent rotations and translations. 
Interestingly and importantly, we find that the molecular configurations having the largest transmission coefficients are generally found 
to be close to those which minimize the system energy. This is due to the fact that when the $\pi$-$\pi$ interaction between the nucleotides 
and the CNTs is large the total energy is generally minimized and the transmission maximized. 

We first determine the optimal distance between the CNTs. The guiding principle for the optimization is that the transmission is expected 
to increase for decreasing CNT separation, but the separation needs to be large enough to allow the translocation of the nucleotides. 
In any hypothetical fully operational device the CNT electrodes' position is fixed as these are embedded in the nanopore, so that we 
have to find a separation which is optimal for all the four nucleotides. In particular the electrodes gap should be wide enough to allow 
also the non-planar methyl group on thymine to pass through. In order to determine the optimal distance between the CNT electrodes we then 
fix the molecule in the lateral position (in the $x$-$y$ plane) that yields the largest transmission at a CNT separation of 6.6~\AA\ and then change 
the separation between the electrodes, always keeping the molecule in the middle of the gap. 
%
\begin{figure}
  \center
  \includegraphics*[width=8.5cm,clip=true]{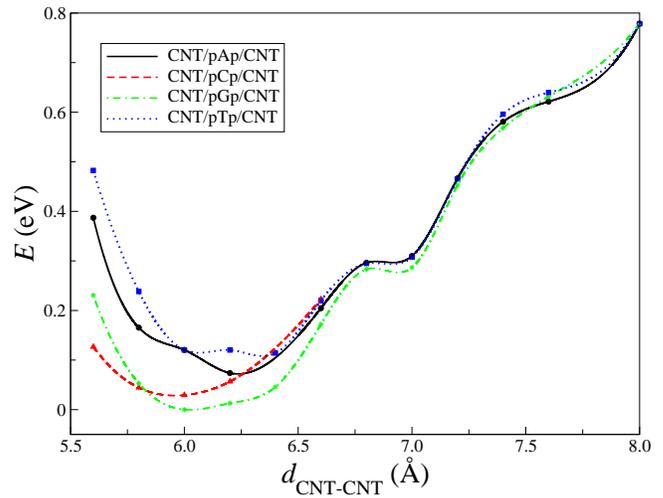}
  \caption{Total energy as function of the distance between the CNT electrodes, $d_\mathrm{CNT-CNT}$, for the
           CNT/pXp/CNT junctions (X=A, C, G, and T). The energy zero is taken at the lowest energy calculated.}
  \label{Fig:EvsCNTDistance}
\end{figure}

We examine first the junction total energy as function of the separation between the CNT caps (see Fig.~\ref{Fig:EvsCNTDistance}). The 
smallest possible CNT electrodes gaps can be seen to be approximately $5.8$~\AA, since there is a sharp increase of the energy for 
closer separations regardless of the nucleotide type. Considering that the optimal distance may be underestimated by DFT calculations 
in which the van der Waals interactions (which tends to be repulsive at short distances) are not included, and that we should use a 
common distance for all four CNT/pXp/CNT junctions, in all subsequent calculations we assume that the device has been set up with 
a distance of $6.4$~\AA\ between the two CNT electrodes. For this electrodes distance the molecular configurations of the four 
CNT/pXp/CNT junctions yielding the maximum transmission at \ef\ are summarized in Table~\ref{Tab:PreferredConfig} in terms of 
the geometrical offset relative to their initial configurations. We denote these configurations as the {\it optimal configurations}.
%
\begin{table}
    \caption{Optimal molecular configurations, which lead to maximal transmission at \ef, 
             for the four CNT/pXp/CNT junctions (X=A, C, G, and T), relatively to the initial 
             configurations. $\theta$ is the right-hand rotation angle of the nucleobases about the 
             $z$-axis. The distance between the CNTs for optimal configurations is chosen to be commonly $6.4$~\AA.}
    \begin{center}
       {\small
       \begin{tabular}{p{100pt}p{30pt}p{30pt}p{30pt}p{30pt}}\hline
                                   & pAp           & pCp           & pGp           & pTp   \\ \hline
            $\theta$                 & $14.32^\circ$ & $12.28^\circ$ & $14.28^\circ$ & $11.08^\circ$\\
            Offset $y$(\AA)        & $-0.21$       & $-0.14$       & $-0.09$       & $ 0.03$      \\
            Offset $x$(\AA, peak 1)& $-0.05$       & $-2.13$       & $-0.58$       & $-1.88$      \\
            Offset $x$(\AA, peak 2)& $--$          & $ 1.01$       & $--$          & $ 1.19$      \\
         \hline
       \end{tabular}}
    \end{center}
    \label{Tab:PreferredConfig}
\end{table}

\begin{figure}
  \center
  \includegraphics*[width=8.5cm,clip=true]{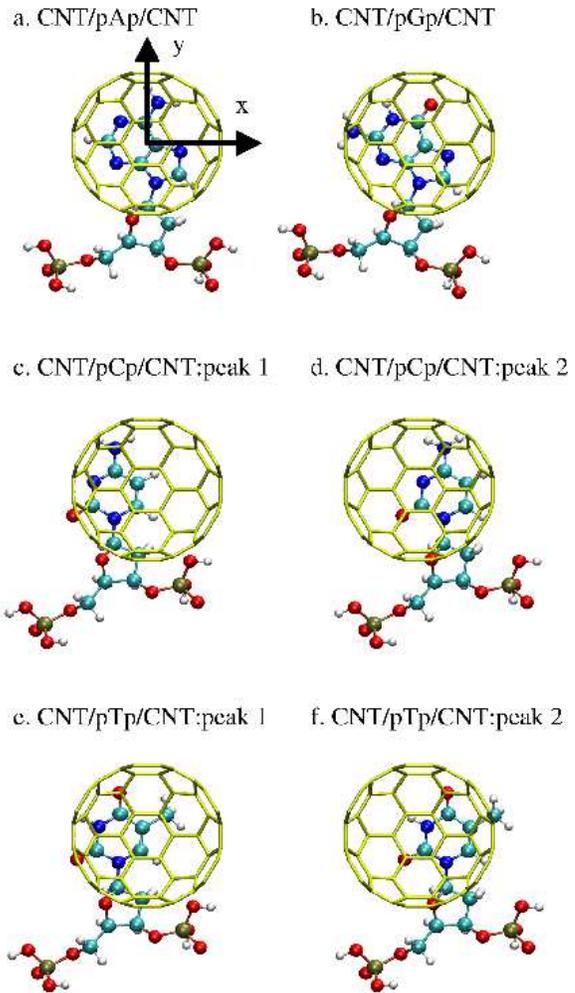}
  \caption{Optimal molecular configurations for the CNT/pXp/CNT junctions (X=A, C, G, and T). For pCp and pTp the 
  two equivalent configurations are both shown.}
  \label{Fig:PreferredConfig}
\end{figure}
An interesting result of our search for the optimal configurations is that pCp and pTp turn out to have two configurations where 
the transmission coefficient peaks (see figure~\ref{Fig:PreferredConfig} for a graphic representation of all the optimal configurations). 
In these the smaller pyrimidine bases of the cytosine and thymine are located at either side of the CNT electrodes along the $x$-axis. 
In contrast, pAp and pGp appear to have only one major peak, and the configuration is such that the larger purine bases of adenine 
and guanine are located in the center of electrodes gap. This difference between purine and pyrimidine bases results in distinct profiles 
of the transmission coefficient along the ssDNA translocation direction (see later in Section~\ref{sec:TalongTranslocation}). In the 
calculations that will follow we will consider always the configuration responsible for the first of the transmission peaks for the 
CNT/pCp/CNT and CNT/pTp/CNT junctions.

\subsection{\label{sec:ElectronicStructure}Electronic structure and zero-bias transmission at the optimal configurations}

\begin{figure}
  \center
  \includegraphics*[width=9cm,clip=true]{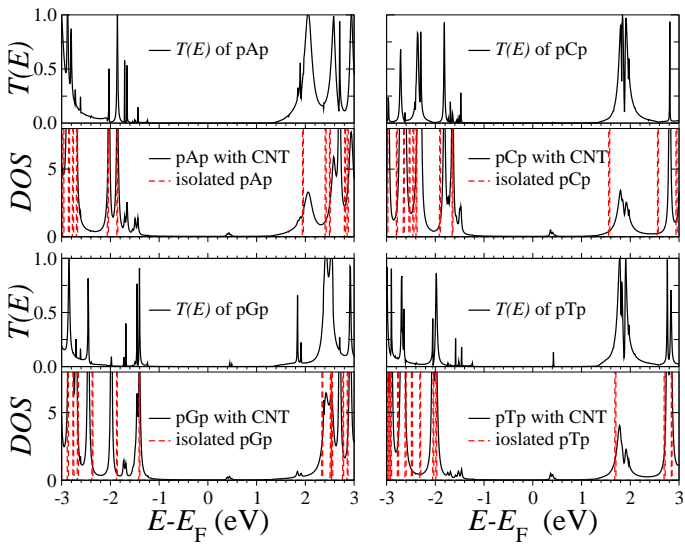}
  \caption{Zero-bias transmission coefficient as function of energy and corresponding density of states (DOS)
           projected onto the pXp molecules for the four CNT/pXp/CNT junctions. In the panels showing the DOS
           we also present, as dashed vertical lines, the DOS of the pXp molecules in the gas phase.}
  \label{Fig:Preferred-PDOS-Tat0V-pXp}
\end{figure}

In order to understand the details of the electron transfer across the CNT/pXp/CNT (X=A, C, G and T) junctions, we now analyze the 
relation between the transmission and the electronic structure of the devices at their optimal configurations. In 
Fig.~\ref{Fig:Preferred-PDOS-Tat0V-pXp} we plot the zero-bias transmission coefficient and the DOS of the pXp molecules 
forming the CNT/pXp/CNT junctions. The vertical dashed lines indicate the energies of the eigenvalues of the isolated pXp 
nucleotides, i.e. they correspond to the DOS of the isolated molecules. From the figure it is clear that the eigenvalues
of the isolated molecules align well with the peaks in the DOS of the corresponding junction (note that a global shift is applied
in order to align the highest occupied molecular orbital -HOMO- of the molecule in the gas phase to that in the junciton). 
This is indicative of the fact that the molecule-electrode interaction does not involve charging, so that the molecule spectrum is 
not distorted in any significant way. However the DOS of the junctions reveals a considerable level broadening, suggesting that there 
is a substantial overlap between the molecular orbitals of the nucleotides and the extended states of the electrodes. 
%

This strong overlap is responsible for the high transmission and Fig.~\ref{Fig:Preferred-PDOS-Tat0V-pXp} reveals a one to one
correspondence between the molecule DOS and the peaks in the transmission functions. In particular for all the four junctions the 
HOMO and the lowest unoccupied molecular orbital (LUMO), both located on pXp molecules, are far away from the electrodes' 
Fermi level. As such the tunneling transmission region around \ef\ corresponds to the molecule HOMO-LUMO gap. A summary 
of the energy position of both the HOMO and the LUMO of the four junctions is presented in Table~\ref{Tab:pXp-HOLU}. 
%
\begin{table}
    \caption{Energy position of the HOMO and the LUMO of the pXp molecules in their corresponding CNT/pXp/CNT junctions 
    (X=A, C, G, and T).}
    \begin{center}
       {\small
       \begin{tabular}{p{50pt}p{40pt}p{40pt}p{40pt}p{40pt}}\hline
                       & pAp        &  pCp        & pGp         & pTp        \\ \hline
            HO (eV)   & $-1.86$    &  $-1.65$    &  $-1.40$    & $ -1.98$   \\
            LU (eV)   & $ 2.06$    &  $ 1.81$    &  $ 2.42$    & $  1.78$   \\ \hline
         \hline
       \end{tabular}}
    \end{center}
    \label{Tab:pXp-HOLU}
\end{table}

\begin{figure}
   \center
   \includegraphics*[width=9cm,clip=true]{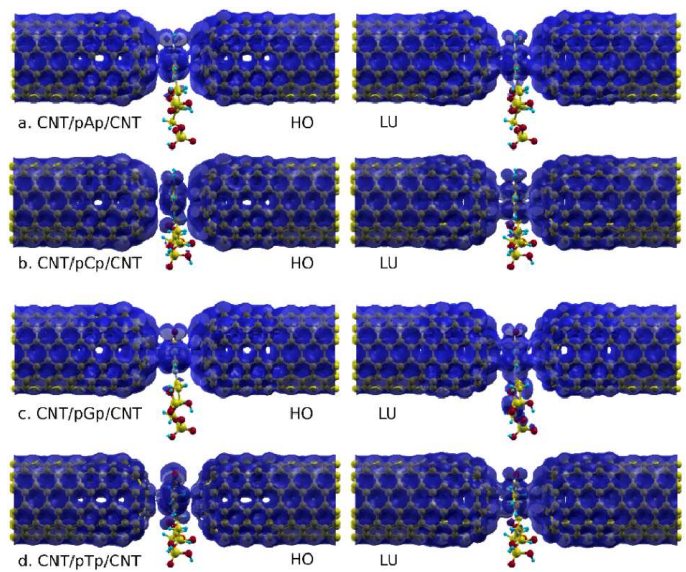}
   \caption{Local density of state (LDOS) isosurfaces of the HOMO and LUMO of the CNT/pXp/CNT junctions at zero-bias. The
   geometries corresponds to the optimal configurations.}
   \label{Fig:Preferred-HOLU}
 \end{figure}

At the optimal configurations (see Fig.~\ref{Fig:PreferredConfig}) the pXp molecules are strongly coupled to the CNT electrodes, 
through the optimal $\pi$-$\pi$ coupling of the nucleobases with the CNT end caps. In order to visualize such a $\pi$-$\pi$ 
bond, we calculate the real space local density of states (LDOS) associated to the charge density of both the HOMO and the LUMO 
of the CNT/pXp/CNT junctions. For the HOMO (LUMO) we integrate the charge density over an energy window $45$~meV ($50$~meV) 
wide around the associated peak in the DOS. The results are shown in Fig.~\ref{Fig:Preferred-HOLU}, where the $\pi$-$\pi$ bonds
are clearly visible. The face-to-face $\pi$-$\pi$ stacking of the six-member rings of the CNT end caps with the nucleobases not only 
makes the optimal configurations energetically favorable, but also enhances the electron tunneling. Since the $\pi$-$\pi$ interaction 
disappears when the nucleobases are outside the central region in between the electrodes, one may expect that the distribution of 
transmission coefficients (and hence the currents) should peak when the nucleotides are inside such a region and that the high
transmission configurations should be rather stable.
%

\begin{figure}
  \center
  \includegraphics*[width=9cm,clip=true]{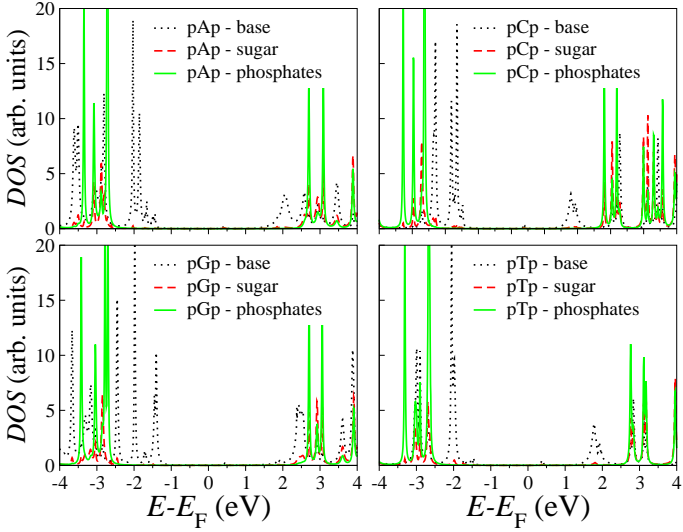}
  \caption{Density of states (DOS) projected onto the base, the sugar and the phosphates of the various pXp molecules in the CNT/pXp/CNT junctions,
   calculated at zero-bias at the optimal configurations.}
  \label{Fig:Preferred-PDOS-pXp-Components}
\end{figure}

Importantly the $\pi$-$\pi$ interaction of the pXp (X=A, C, G, and T) molecules with the CNTs involves only the nucleobases. 
This means that the sugar ring and two nearest phosphate groups in pXp do not play any significant r\^ole either in the bonding or in the 
electron transport. A demonstration of this feature is provided in Fig.~\ref{Fig:Preferred-PDOS-pXp-Components}, where we present the
DOS projected respectively onto the nuclebase, the sugar and the phosphate groups. From the figure it is clear that both the HOMO and the LUMO 
are almost entirely associated to the nuclebases, while both the sugar and the phosphate groups contribute only to molecular levels far way from the 
Fermi level. This is a rather important aspect of the electronic structure of the DNA/CNT system investigated here, namely the fact that the 
non-sequence specific features of the DNA spectrum are away from the electrodes Fermi level and therefore are not expected to 
contribute significantly to the transport at least at relatively moderate voltages.
%

\begin{figure}
  \center
  \includegraphics*[width=3.25in,clip=true]{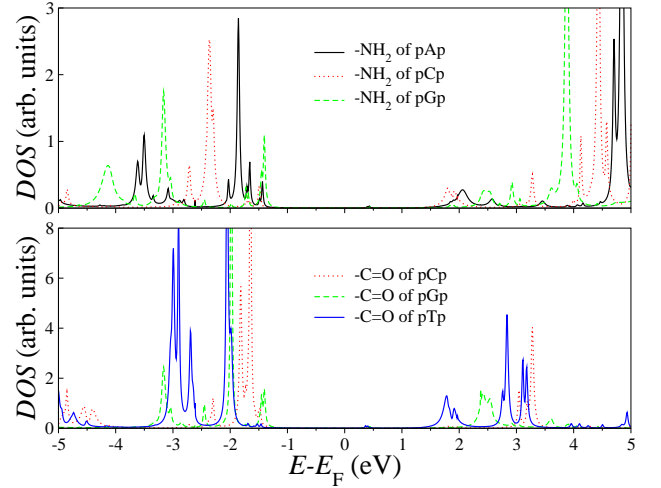}
  \caption{Density of states (DOS) projected onto the amine and carbonyl groups of the various pXp molecules in the CNT/pXp/CNT junctions,
   calculated at zero-bias at the optimal configurations.}
  \label{Fig:Preferred-PDOS-pXp-NH2CO}
\end{figure}

Finally we wish to investigate further which functional part of the nucleotides contributes the most to the $\pi$-$\pi$ bonding by looking at the DOS
projected either on the amine (NH$_2$) or the carbonyl (C=O) group (see Fig.~\ref{Fig:Preferred-PDOS-pXp-NH2CO}). Note that pAp, pCp and pGp 
have one primary amine group, pCp and pGp have one carbonyl and finally pTp has two carbonyls. From Fig.~\ref{Fig:Preferred-PDOS-pXp-NH2CO}
we note that for pAp, the primary amine group contributes the most to the HOMO, with a major peak in the DOS at around $-1.86$~eV and minor peaks at around 
$-1.66$~eV and $-1.44$~eV. These form because of the coupling to the CNT electrodes. The presence of such minor peaks helps to explain why at zero-bias pAp has the largest 
transmission coefficient despite its major HOMO peak lies below those of pCp and pGp (Table~\ref{Tab:pXp-HOLU}). In contrast for pCp, the carbonyl group 
contributes more than the primary amine one to the HOMO, while for pGp these two groups contribute similarly to their HOMO. The main peaks of the 
carbonyl groups appear to be similar for pGp and pTp (around $-2$~eV) but these are located at higher energies for pCp (around $-1.65$~eV). The 
different positioning and strength of the electronic states located on these functional groups are important, since they affect the tunneling signals for the different 
nucleobases. Their characteristic signals may be augmented by functionalizing the electrodes with suitable chemical agents in order to amplify the 
electronic contrast between the different nucelobases\citep{Ohshiro2006, He2007, He2008}. 
%

\subsection{$I$-$V$ characteristics at the optimal configurations}
\label{sec:TransportProperties}

\begin{figure}
  \center 
  \includegraphics*[width=8cm,clip=true]{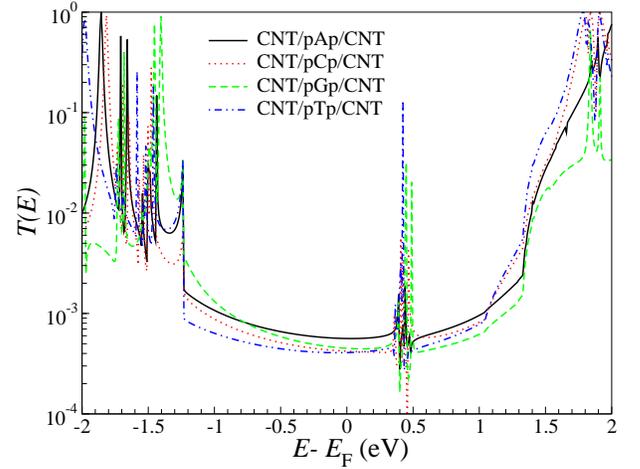}
  \caption{Zero-bias transmission coefficient as a function of energy for CNT/pXp/CNT (X=A, C, G, and T) junctions at the optimal configurations.}
  \label{Fig:Preferred-Tat0V}
\end{figure}

We now investigate the transport properties of the pXp molecules (X=A, C, G, and T) by assuming that these are ideally measured at the optimal configurations. 
Fig.~\ref{Fig:Preferred-Tat0V} shows the zero-bias transmission coefficient, $T(E)$, plotted on a logarithmic scale. As noted previously, the peaks in transmission 
at around $0.5$~eV are due to localized surface states belonging to the CNT close-ends. These are unaffected by the pXp configuration and
they contribute little to the electron current as bias is applied (for a discussion on the resonant transport properties of surface localized state see reference 
[\onlinecite{IvanTMR}]).
%
At the Fermi level the relative order of magnitude of the transmission coefficient is now $T_\mathrm{pAp}$ $>$ $T_\mathrm{pGp}$ $>$ $T_\mathrm{pCp}$ $>$ $T_\mathrm{pTp}$,
i.e. it is the same order that we have found for the initial configurations (see Fig.~\ref{Fig:InitialConfig-Tat0V}). However, if one now looks at energies away from $E_\mathrm{F}$
a few changes are notable. For instance at $-0.77$~eV (energies are always taken from $E_\mathrm{F}$) the transmission coefficient of pGp surpasses that of pAp and the 
order becomes $T_\mathrm{pGp}$ $>$ $T_\mathrm{pTp}$ $>$ $T_\mathrm{pAp}$ $>$ $T_\mathrm{pCp}$. In contrast for positive energies $T_\mathrm{pTp}$ and 
$T_\mathrm{pCp}$ first surpass $T_\mathrm{pGp}$ at $0.5$~eV and then $T_\mathrm{pAp}$ at $1.07$~eV. In the region $[1.1, 1.8]$~eV, the order is now  
$T_\mathrm{pTp}$ $>$ $T_\mathrm{pCp}$ $>$ $T_\mathrm{pAp}$ $>$ $T_\mathrm{pGp}$, which roughly reflects the energy order of the various LUMOs
(Table~\ref{Tab:pXp-HOLU}). Such a rather sensitive energy dependence of the order of the transmission coefficients for the various molecules reflects the fact
that the nucleotides are electronically rather similar. However, we will show that the $I$-$V$ curves offer a reasonable voltage range where the molecules can be
distinguished. 

\begin{figure}
  \center
  \includegraphics*[width=8cm,clip=true]{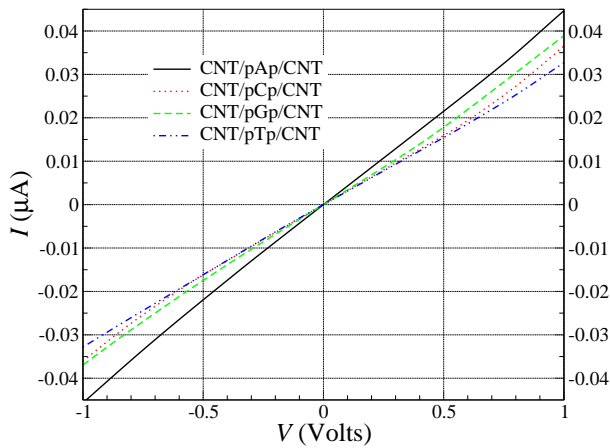}
  \caption{$I$-$V$ characteristics for the CNT/pXp/CNT (X=A, C, G and T) junctions at the optimal configurations in the [-1, 1]~Volt bias range.}
  \label{Fig:Preferred-IV-LowBias}
\end{figure}

The calculated $I$-$V$ curves in the moderate bias range [-1, 1]~Volt are presented in Fig.\ref{Fig:Preferred-IV-LowBias}. In this particular bias window the transport 
is well within the tunneling regime, since there are no peaks in the transmission attributable to any pXp molecular orbitals (see Fig. \ref{Fig:Preferred-Tat0V}). Importantly 
in this moderate bias region the order of the currents amplitude is the same as that of the transmission coefficients at the Fermi level, namely 
$I_{\textrm{pAp}} > I_{\textrm{pGp}} > I_{\textrm{pCp}} > I_{\textrm{pTp}}$. This is persistent throughout the entire bias values investigated, suggesting that a current
measurement in this range should be able to distinguish between the nucleotides. Furthermore, the current curves in Fig.\ref{Fig:Preferred-IV-LowBias} are well spaced 
when the bias is between $0.7$~V and $1$~V, a bias interval which therefore emerges as the ideal probing voltage range. We note that recently Tsutsui \et reported a similar trend, 
namely $I_{\textrm{G}} > I_{\textrm{C}} > I_{\textrm{T}}$, for the peak currents of single nucleotides measured by two gold nanoelectrodes at a constant dc bias of 
$0.75$~V \citep{Tsutsui2010}.
%

\begin{figure}
  \center
  \includegraphics*[width=8cm,clip=true]{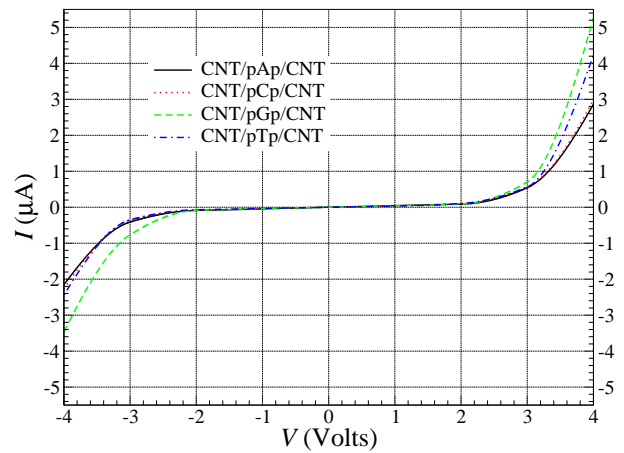}
  \caption{$I$-$V$ characteristics for the CNT/pXp/CNT (X=A, C, G and T) junctions at the optimal configurations in the [-4, 4]~Volt bias range.}
  \label{Fig:Preferred-IV-FullRange}
\end{figure}

In Fig.~\ref{Fig:Preferred-IV-FullRange} the $I$-$V$ curves for the same configurations are shown over an extended bias window ranging now from $-4$~V to $4$~V. 
From the figure we can clearly identify the transition from tunneling to a resonant transport regime at about $\pm 3$~V for all nucleotides. This is the voltage which 
coincides with a drastic increase of the current. For voltages larger than $3.5$~V and smaller than $-3.5$~V, the current signals associated to the different nucleotides
have a steady order, namely $I_{\textrm{pGp}} > I_{\textrm{pTp}} > I_{\textrm{pCp}} \sim I_{\textrm{pAp}}$ (Fig.~\ref{Fig:Preferred-IV-FullRange}). Also notable is the 
fact that the $I$-$V$ characteristics are approximately symmetric with respect to the bias polarity, since the nucleotides are placed in the middle of the gap between 
the CNT electrodes. 
%

\subsection{\label{sec:TalongTranslocation}Transmission profile of nucleotides along the translocation path}

In the discussion presented in the previous section, each of the four nucleotides is placed at the optimal configuration in the corresponding 
CNT/pXp/CNT (X=A, C, G, and T) junction. We now examine the evolution of the transport properties of the nucleotides when they translocate through 
the gap between the CNT electrodes along the $x$-axis. This is precisely the motion that the nucleotides will undergo in a nanopore experiment 
(Fig.~\ref{Fig:CNT-Device}). For simplicity, the pXp molecules, which have limited conformational flexibility near the $6.4$~\AA-wide electrode gap, 
are assumed to be rigid. Their positions are measured as the distance between the nucleotide center of mass and the mid-point of CNT electrodes.

\begin{figure}
  \center
  \includegraphics*[width=8cm,clip=true]{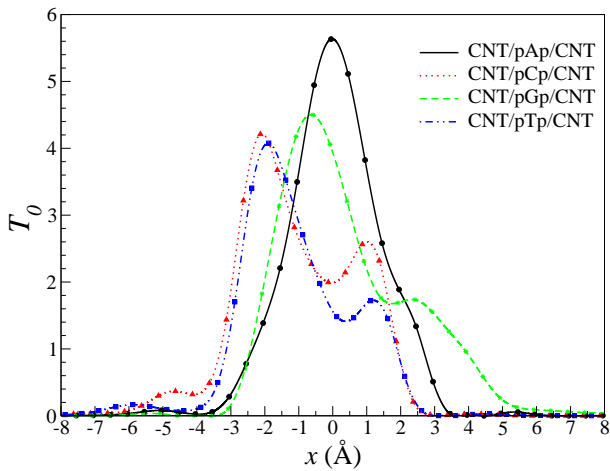}
  \caption{Transmission coefficient at the Fermi level along the DNA translocation path.}
  \label{Fig:TranslocationX-ACGT-Tat0V}
\end{figure}

Fig.~\ref{Fig:TranslocationX-ACGT-Tat0V} shows the profiles of the transmission coefficient at the Fermi level, $T(E_\mathrm{F})=T_{0}$, along the nucleotides' 
translocation paths, with the peaks appearing at the optimal configurations [see Table~\ref{Tab:PreferredConfig}]. As already mentioned in Section~\ref{sec:SearchConfigs}, 
one can clearly see that the profiles of pCp and pTp are distinguishable from those of pAp and pGp by having one more pronounced transmission peak. 
The double peak structure arises from the two nearly degenerate optimal configurations, that both pAp and pGp possess  because of their different
symmetry. Since both the cytosine and the thymine base can be flipped before entering the electrode gap, the order and the relative magnitude of these 
two peaks can be reversed. In addition we also note that the profile of pGp appears to have a broad shoulder, indicative of a second high transmission configuration
along the translocation path. From Fig.~\ref{Fig:TranslocationX-ACGT-Tat0V} we also observe that the transmission coefficient decays sharply once the nucleotides 
is moved outside the electrodes gap. Notably the radius of the (6,6) CNT is about $4.1$~\AA\ while the distance between adjacent nucleotides in a stretched ssDNA 
is about $7$~\AA. As the tails of the transmission coefficient curves are quite small when $x<-3.5$~\AA\, or $x>3.5$~\AA~(i.e., when the nucleobase is outside of 
electrodes gap), we conclude that having nucleotides in a sequence should not affect much the reading of the tunneling current of a single nucleotide. 

We next integrate $T_0$ along the translocation path to obtain the aggregate transmission coefficient (denoted as $AT$). Our results are summarized in 
Table~\ref{Tab:TranslocationX-AggregateT}. $AT$ over the path $-3.5<x<3.5$~\AA\ accounts for 99~\%, 94~\%, 92~\% and 96~\% of the $AT$ calculated over the 
more extended path  $-8<x<8$~\AA, respectively for pAp, pCp, pGp and pTp. This is a further indication that most of the current is collected when the nucleobase 
is within the CNT region and that adjacent nucleobases are far enough not to affect the measurement. Similar conclusions have been reached\cite{Zwolak2005}
for Au electrodes at separations smaller than 1~nm.

The aggregate transmission coefficient represents the time-average low-bias conductance over the translocation of each nucleotide, once we assume that the translocation 
occurs at a constant velocity. Interestingly, we find a well spaced order in the $AT$, namely $AT_{\textrm{pAp}} > AT_{\textrm{pGp}} > AT_{\textrm{pCp}} > AT_{\textrm{pTp}}$,  
both when the $AT$ is calculated within $\pm 3.5$~\AA\ or $\pm 8$~\AA\ from the electrode gap center. Notably the order of the integrated transmission corresponds to the 
order found for the transmission at the optimal positions, which indeed dominate the transport during the translocation. More importantly this fact indicates that the 
pAp$>$pGp$>$pCp$>$pTp order is recurrent in most of the transport measurements we have described and therefore it is a rather robust result of the geometry 
investigated here.
%

\begin{table}
    \caption{Aggregate transmission coefficient at the Fermi level along the DNA translocation path centered in the middle plane between the CNT electrodes (unit: $10^{-4}$~\AA).}
    \begin{center}
       {\small
       \begin{tabular}{p{60pt}p{40pt}p{40pt}p{40pt}p{40pt}}\hline
                                       & pAp      &  pCp       & pGp      & pTp  \\ \hline
          $\int_{-3.5}^{3.5} T_{0}\,dx$  & $17.03$  &  $14.06$   & $15.81$  & $12.30$  \\
          $\int_{-8.0}^{8.0} T_{0}\,dx$  & $17.24$  &  $14.93$   & $17.19$  & $12.82$  \\
         \hline
       \end{tabular}}
    \end{center}
    \label{Tab:TranslocationX-AggregateT}
\end{table}

\subsection{\label{sec:ConfigutationEffect}Effect of molecular configurations on the transport properties}

\begin{figure}
  \center
  \includegraphics*[width=8cm,clip=true]{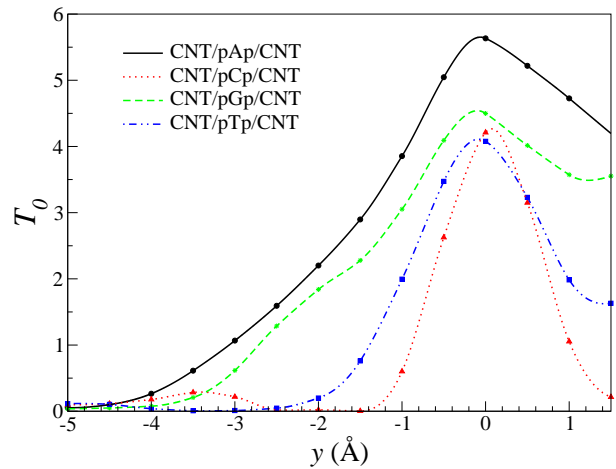}
  \caption{Transmission coefficient at Fermi level for the CNT/pXp/CNT (X=A, C, G, and T) junctions as the nucleotides move along the $y$-axis.}
  \label{Fig:Factors-VariationInY}
\end{figure}

In the previous discussion about the transmission profiles, we have assumed that the nucleotides translocate ideally along the $x$ direction without any lateral displacement within
the electrodes gap (along the $y$-axis). We now examine how $T_{0}$ varies when the nucleobases position fluctuates along the $y$ direction, by using the nucleobase center
of mass of the relative optimal configuration as origin, $y=0$. As it can be seen in Fig.~\ref{Fig:Factors-VariationInY}, $T_{0}$ decays quickly when the nucleobases move downwards away 
from the electrodes (negative $y$). This is similar to what we have observed in Fig.~\ref{Fig:TranslocationX-ACGT-Tat0V} for a longitudinal displacement. Note that there is not
much room for positive $y$ displacement now as this will need the non-planar sugar ring and phosphate groups to enter the narrow electrodes gap. As such we stop the
calculations at $y=1.5$~\AA, which corresponds to the largest displacement possible along that direction.
%

\begin{figure}
  \center
  \includegraphics*[width=8cm,clip=true]{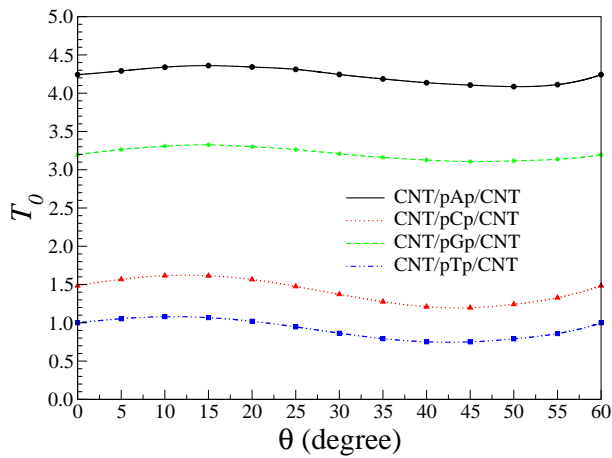}
  \caption{Transmission coefficient calculated at the Fermi level for the CNT/pXp/CNT (X=A, C, G, and T) junctions as a function of the rotation of angle of the pXp molecules about the $z$-axis (vertical).}
  \label{Fig:Factors-CNT-Rotation}
\end{figure}

Next we consider the effect of rotations of the pXp molecules about the $z$-axis. As shown in Fig.~\ref{Fig:PreferredConfig}, at the optimal configurations, the six-member ring
of the end cap in each of the CNT electrodes is parallel to the plane of the nucleobases. Moreover, the cap ring tends to align with the pyrimidine rings of the purine bases 
(adenine and guanine) so that the corresponding hexagon edges are parallel to each other. In contrast the edges are staggered with the pyrimidine rings of cytosine and thymine so 
that one of its vertex atoms faces the center of the pyrimidine. One may then expect that rotations may produce significant change in the low-bias conductance. This is however not
the case. In fact by plotting $T_0$ as a function of the rotation angle, $\theta$, (Fig.~\ref{Fig:Factors-CNT-Rotation}) we find only small changes, and most importantly also that the 
order of the amplitudes of the conductance of the various nucleobases does not change. This essentially means that rotations about the optimal configurations will not undermine our 
ability to identify the nucleobases.
%

\begin{figure}
  \center
  \includegraphics*[width=8cm,clip=true]{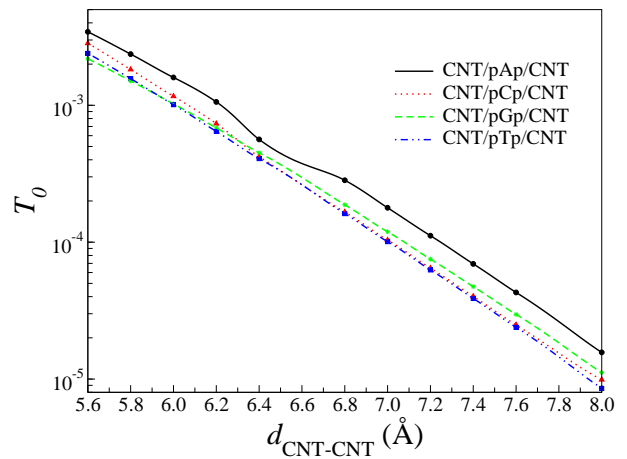}
  \caption{Transmission coefficient at Fermi level for the CNT/pXp/CNT (X=A, C, G, and T) junctions as a function of the electrode separation, $d_\mathrm{CNT-CNT}$. Note that $T_0$ decays exponentially when $d_\mathrm{CNT-CNT}$
          increases.}
  \label{Fig:Factors-CNT-Distance}
\end{figure}

Finally, we investigate the effects that the gap size between the CNT electrodes has on the transport by plotting $T_{0}$ as function of the electrode distance, $d_\mathrm{CNT-CNT}$ (Fig.~\ref{Fig:Factors-CNT-Distance}).
In this case we maintain the position of the plane containing the pXp molecules fixed at the midpoint between the electrodes. As expected from a tunneling process we find that the transmission 
coefficient decreases exponentially with increasing $d_\mathrm{CNT-CNT}$. The figure~\ref{Fig:Factors-CNT-Distance} also shows that the exponential decay rate $\alpha$, defined as the slope of the 
$\ln T (d)$ curves, is rather similar among the nucleotides. Interestingly $T_0$ for pAp is always well separated from that of the other nucleotides for all the electrode gap distances investigated. In contrast 
the relative order between pCp, pGp and pTp depends on $d_\mathrm{CNT-CNT}$. It then emerges that it is absolutely crucial to maintain the electrode separation as stable as possible during
a measurement and also that this should be the smallest possible, so that configurational fluctuations will be highly suppressed.
%

\subsection{\label{sec:SequencingProtocol}DNA sequencing protocol}

With all the results of the previous sections in mind we can now formulate a possible protocol for reading the nucleotide sequence. The proposed device functions by translocating ssDNA through a nanopore
and simultaneously by measuring the transverse current between two closely spaced CNT nanotubes [(6,6) close-ended CNTs at 6.4~\AA\ separation in our case]. The DNA strand can be pulled through 
the nanopore mechanically by an optical or magnetic tweezer~\cite{Keyser2006, Peng2009} with an approximately constant velocity. Typical velocities for optical tweezers have been 
reported~\cite{Keyser2006} at around 30~nm$\cdot$s$^{-1}$. This means that one need a time resolution in the kHz range in order to obtain a spatial resolution of 0.1~\AA, which is certainly enough for 
detecting accurately the transmission profiles of the nucleotides shown in Fig.~\ref{Fig:TranslocationX-ACGT-Tat0V}. Note that the currents involved are of the order of 10~nA and we need an amplitude 
resolution of the order of 1~nA or less (depending on the bias) for typical resistances of the order of 10~M$\Omega$. These are rather standard demands for conventional electronics. 

In contrast if the ssDNA translocation is electrophoretically driven, then the velocity may fluctuate considerably so that the recognition of the current profiles may become difficult. 
Furthermore the typical translocation velocities are much larger than those achievable by mechanical manipulations. For instance early studies\cite{Fologea2005} reported speeds of the order of 
10$^6$~nm$\cdot$s$^{-1}$, which require GHz time-resolution for a sub-\AA\ spatial resolution. More recently a substantial slowing down has been achieved by gate modulation of the nanopore wall surface 
charges, with reported velocities of 55~$\mu$m$\cdot$s$^{-1}$. This translates in a 5~MHz time resolution for 0.1~\AA\ spatial resolution. Clearly GHz frequencies are out of the reach, unless the typical 
resistances are in the k$\Omega$ range, while MHz resolution may be possible. In any case we believe that for a usable device setup electrophoretically driven translocation
is not suitable at present, at least if the signal is made of the current profiles during translocation.

Further information however can be extracted by simply taking the integral over time of each current peak and/or successive peaks. These are acquired during the translocation of an entire 
nucleobase across the gap between the electrodes. As such the time-resolution required is now about two orders of magnitude larger than that needed for resolving the profiles. This means 
operating the device at $\sim$100~kHz frequencies for slow electrophoretically driven translocations. 

Finally an alternative quantity to measure is, as proposed first by Zwolak \et \citep{Zwolak2005}, the statistical distribution of the current over time. This is a measure probably achievable only
with the tweezers setup and essentially consists in holding the ssDNA at a fixed position and then, after the system has thermally equilibrated inside the nanopore, in measuring the current. 
We expect that the details of the current distributions may be device-dependent, but the peak currents of the nucleobases should maintain the order 
$I_{\textrm{pAp}} > I_{\textrm{pGp}} > I_{\textrm{pCp}} > I_{\textrm{pTp}}$, at least in the suitable low bias range around $[0.7, 1]$~V. 

\begin{figure}
  \center
  \includegraphics*[width=8.5cm,clip=true]{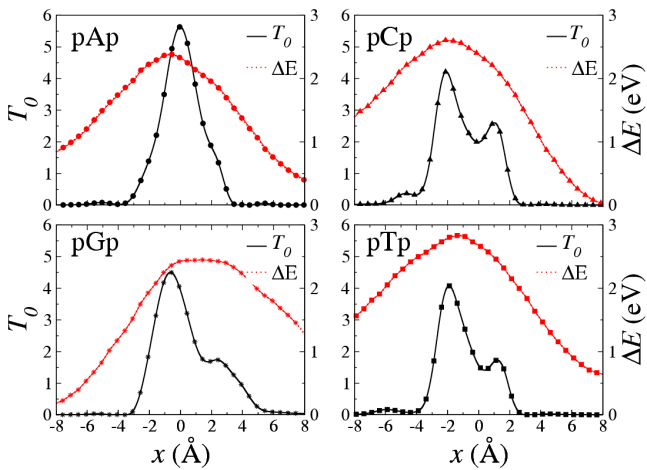}
  \caption{Transmission coefficient at the Fermi level against the interaction energy for the CNT/pXp/CNT (X=A, C, G, and T) junctions along DNA translocation path.}
  \label{Fig:TranslocationX-ACGT-TvsE}
\end{figure}

The statistical distributions of the transmission coefficient/current of the CNT/pXp/CNT junctions are yet to be calculated for our device (Fig.~\ref{Fig:CNT-Device}). However,
some insights can be obtained by plotting the $T_0$ together with the interaction energy between the pXp molecules and the CNT electrodes along the
translocation path. The interaction energy is defined as $\Delta E = E_{\textrm{pXp}}+E_{\textrm{CNT}}-E_{\textrm{CNT/pXp/CNT}}$, where $E_\alpha$
is the DFT total energy of the $\alpha$ system ($\alpha=$~pXp, CNT and CNT/pXp/CNT). Our results are shown in Fig.~\ref{Fig:TranslocationX-ACGT-TvsE}. The calculated 
interaction energies are one order of magnitude larger than that for typical hydrogen bonding ($0.1-0.4$~eV). This is due to the strong $\pi$-$\pi$ interaction found in our device 
at the optimal configurations. Interestingly, we find that the position of the maximum transmission corresponds approximately to the position displaying the largest interaction energy.
As such, one may expect that when a nucleobase passes through the device, it will tend to remain between the electrodes and parallel to the end caps of CNTs. In other words
the same interaction responsible for the relatively large currents and for the suppression of the geometry fluctuations also acts to slow down the DNA translocation
across the nanopore. 

\section{\label{sec:Conclusions}Conclusions}

In summary we have proposed a device for high throughput DNA nanopore sequencing, based on reading the transverse current across the nucleobases of ssDNA as it passes between 
two closely spaced single-walled close-ended (6,6) CNTs. The device operation has been investigated theoretically with state of the art density functional theory combined with non-equilibrium
quantum transport. In general the electron transport through the various nucleobases can be roughly separated into a tunneling regime at low bias and a resonant transport 
one at higher bias. We have identified the optimal position of the various nucleotides yielding the highest zero-bias conductance and used them to analyze the various parameters affecting
the device. In particular we have focussed our attention on the dependence of the electron transmission on the longitudinal position of the nucleotides with respect to the center of the
electrodes, on possible lateral displacements and on the electrode separation. 

The current/conductance profiles of the nucleotides along the DNA translocation path reveal that the pyrimidine bases (cytosine and thymine) have a two peaks structure, indicative
of two high transmission geometries. The same is not found for the other two bases. Importantly we have found that the peak tunneling currents of the nucleobases have the following
order $I_{\textrm{pAp}} > I_{\textrm{pGp}} > I_{\textrm{pCp}} > I_{\textrm{pTp}}$ at low bias and that the same order is maintained for the aggregate transmission coefficients. 
Furthermore we have demonstrated that at this close electrode separation the current drops drastically as the nucleotides exit the electrodes region, so that well separated
current signals should be detected during the translocation. 

Finally we have proposed a measurement protocol for rapid DNA sequencing. This is presently suitable for tweezer driven translocation and consists of three independent measurements.
The first involves the pattern recognition of the current profile during the translocation of the nucleobases. The second relies on the fact that the aggregate transmission coefficient (and hence the 
integral of the current) along the translocation path has a well-defined order. Finally the third concerns the time-distribution of the current for mechanically stabilized junctions, where the
nucleotide to measure is held at a fixed position. 

\begin{acknowledgments}
XC gratefully acknowledges the European Community's Seventh Framework Programme  for financial support (Project: nanoDNAsequencing: NanoTools for Ultra Fast DNA Sequencing; 
Grant agreement no.: 214840). Additional financial support has been provided by (CDP and SS) Science Foundation of Ireland (Grant No. 07/IN.1/I945) and (IR, US and SS) by the King Abdullah 
University of Science and Technology (ACRAB project). Calculations have been performed on the Shaheen supercomputer at the King Abdullah University of Science and Technology.
Illuminating discussion with Nadjib Baadji are kindly acknowledged.
\end{acknowledgments}


\providecommand{\noopsort}[1]{}\providecommand{\singleletter}[1]{#1}%

\end{document}